\documentclass[conference]{IEEEtran}

\usepackage{array}
\usepackage{cite}
\usepackage{amsmath}
\usepackage{bbm}
\usepackage{algorithmic}
\usepackage{xcolor}
\usepackage{url}

\makeatletter
\let\MYcaption\@makecaption
\makeatother
\usepackage[font=footnotesize]{subcaption}
\makeatletter
\let\@makecaption\MYcaption
\makeatother

\ifCLASSINFOpdf
    \usepackage[pdftex]{graphicx}
    \graphicspath{{../pdf/}{../jpeg/}}
    \DeclareGraphicsExtensions{.pdf,.jpeg,.png}
\else
    \usepackage[dvips]{graphicx}
    \graphicspath{{../eps/}}
    \DeclareGraphicsExtensions{.eps}
\fi

\usepackage{tikz}
\usepackage{textcomp}
\usepackage{hyperref}
\usepackage{lipsum}

\newcommand\copyrighttext{%
  \footnotesize \textcopyright 2021 IEEE. Personal use of this material is permitted.
  Permission from IEEE must be obtained for all other uses, in any current or future
  media, including reprinting/republishing this material for advertising or promotional
  purposes, creating new collective works, for resale or redistribution to servers or
  lists, or reuse of any copyrighted component of this work in other works.}
\newcommand\copyrightnotice{%
\begin{tikzpicture}[remember picture,overlay]
\node[anchor=south,yshift=10pt] at (current page.south) {\fbox{\parbox{\dimexpr\textwidth-\fboxsep-\fboxrule\relax}{\copyrighttext}}};
\end{tikzpicture}%
}

\begin{document}
\bstctlcite{IEEEexample:BSTcontrol}

\title{Impacts of Time-of-Use Rate Changes on the Electricity Bills of Commercial Consumers}

\author{\IEEEauthorblockN{Lane D. Smith and Daniel S. Kirschen}
\IEEEauthorblockA{Department of Electrical \& Computer Engineering, University of Washington, Seattle, WA, USA\\
Email: \{ldsmith, kirschen\}@uw.edu}}

\maketitle

\copyrightnotice

\begin{abstract}
Changes in the profile of prices in wholesale electricity markets prompt utilities to redesign their tariffs and adjust their time-of-use periods to ensure a more adequate cost recovery. However, changing the rate structures could adversely affect commercial consumers by increasing their electricity bills and hindering their ability to reduce costs using techniques like net energy metering. As time-of-use periods are adjusted, consumers will need to rely on the flexibility of distributed energy resources to achieve cost reductions. This paper explores the effect that Pacific Gas and Electric Company’s redesigned rates have on the electricity bills of consumers with different demand profiles. Sensitivity analyses are conducted to examine the effect of asset sizing on reducing costs under each tariff.
\end{abstract}

\begin{IEEEkeywords}
Distributed energy resources, electric tariffs, power system economics, valuation
\end{IEEEkeywords}

%
\IEEEpeerreviewmaketitle

\section*{Nomenclature}
\addcontentsline{toc}{section}{Nomenclature}
\begin{IEEEdescription}[\IEEEusemathlabelsep\IEEEsetlabelwidth{$SOC_{max}$}]
    \item[\textit{Sets and Indices}]
    \item[$\mathcal{T}$] Set of 15-minute increments in the month, indexed by $t$.
    \item[$\mathcal{P}$] Set of time-of-use periods, indexed by $p$.
    \item[]
    
    \item[\textit{Asset Variables and Parameters}]
    \item[$P_{pv}(t)$] Power generated by a photovoltaic (PV) array.
    \item[$J(t)$] Battery's state of charge.
    \item[$J_{init}$] Initial state of charge.
    \item[$J_{min}$] Minimum state of charge.
    \item[$J_{max}$] Maximum state of charge.
    \item[$\eta$] Battery round-trip efficiency.
    \item[$P_{cha}(t)$] Battery's charging power.
    \item[$P_{dis}(t)$] Battery's discharging power.
    \item[$BER$] Battery's energy rating.
    \item[$BPR$] Battery's power rating.
    \item[]
    
    \item[\textit{Utility Tariff Variables and Parameters}]
    \item[$D_{max}$] Maximum monthly demand.
    \item[$DR_{max}$] Charge rate for maximum monthly demand.
    \item[$D_{tou}(p)$] Maximum monthly demand during time-of-use period $p$.
    \item[$DR_{tou}(p)$] Charge rate for maximum monthly demand during time-of-use period $p$.
    \item[$D_{base}(t)$] Consumer base demand.
    \item[$D_{net}(t)$] Net consumer demand.
    \item[$D_{net}^{+}(t)$] Consumer imports from the grid (nonnegative).
    \item[$ER(t)$] Charge rate for energy usage.
    \item[$NSR(t)$] Sell rate for net energy metering.

\end{IEEEdescription}

\section{Introduction}
The advent of distributed solar, catalyzed by decreased capital costs and programs such as net energy metering (NEM), has caused a shift in the profile of prices in wholesale electricity markets: prices are lower during times of peak solar generation (\textit{i.e.,} midday) and higher during times of peak demand (\textit{i.e.,} early evening) \cite{hobbs2019}. For electric utilities that offer flat time-of-use (TOU) rates, this change can be problematic, especially if their TOU periods are poorly aligned with the prices observed in the wholesale markets \cite{darghouth2014, felder2014}. In recent years, utilities such as Pacific Gas and Electric Company (PG\&E) have worked to better align their TOU rates with the prices observed on the wholesale markets so as to better ensure cost recovery \cite{b19}. 

However, restructuring these rates can have a significant impact on a consumer's total electricity bill and on its ability to reduce its electricity cost using distributed energy resources (DERs) \cite{darghouth2014, darghouth2016, darghouth2011}. Shifting TOU periods to align more closely with prices on the wholesale market affects consumers differently depending on the shape of their demand profile and on the types and sizes of the DERs they have deployed. For consumers with distributed solar, realigning TOU periods to occur later in the day undoubtedly reduces the benefits of NEM \cite{darghouth2014}. As the ability of programs like NEM to provide cost savings decreases, consumers who seek to reduce their electricity bills will need to make their demand more flexible, perhaps by installing energy storage or implementing price-based demand response.

In this paper, we explore the impact that PG\&E's redesigned tariffs have on the total electricity bills of commercial consumers with different demand profiles. To adequately explore this impact, an optimization model is created that determines the minimum total bill for consumers with a combination of PV and battery energy storage (BES) assets. This model allows the total bills under existing tariffs to be juxtaposed with those under the redesigned rates. Through sensitivity analyses, we are also able to examine how the size of these assets reduces cost under each tariff. 

The rest of this paper is organized as follows. Section \ref{Overview of PGE Tariffs} provides an overview of the PG\&E tariffs explored in this paper. Section \ref{Mathematical Formulation} presents the mathematical formulation used to minimize the electricity bill of a commercial consumer with a combination of PV and BES assets. Section \ref{Case Studies} introduces case studies of two different consumer types that are eligible for the existing and redesigned PG\&E tariffs. The corresponding electricity bills, asset values, and cost sensitivities are discussed for combinations of consumer assets and tariffs. Section \ref{Conclusions} concludes the paper.

\section{Overview of PG\&E Tariffs} \label{Overview of PGE Tariffs}
As of 2020, two main electric tariffs that PG\&E offers its commercial consumers are the E-19 and B-19 rate schedules \cite{e19, b19}. Beginning in November 2019, B-19 rates became available on a voluntary basis, with the intent to replace the longstanding E-19 rates in March 2021. While both tariffs offer TOU rate schedules, the E-19 and B-19 tariffs are significantly different. Many of these differences indicate a shift towards PG\&E's rates becoming more coincident with current wholesale market prices. One difference is that compared with the E-19 rates, the peak and partial-peak rates under B-19 are much later in the day. For instance, peak rates (\textit{i.e.,} the most expensive rates) under B-19 occur from 4pm to 9pm, rather than between 12pm and 6pm for E-19. Depending on their demand profile, this shift has obvious cost implications for consumers. Consumers participating in NEM are also affected, as the higher sell rates are less aligned with times of peak solar generation. The B-19 tariff also includes a new `super off-peak' rate, which occurs from 9am to 2pm during spring months and features the lowest energy prices. While potentially beneficial for consumers with midday demand peaks, these rates can hurt the sell rate for NEM participants. Finally, unlike the E-19 tariff, where the off-peak rates (\textit{i.e.,} the cheapest rates) were effective all day on weekends and holidays, the B-19 tariff implements TOU pricing every day of the week. Consumers who may have once found price relief on weekends will need to incorporate more intra-day flexibility to find similar relief under B-19 rates.

In this paper, we consider multiple rate options available under the E-19 and B-19 tariffs. Both tariffs offer base TOU rates and Option R for Solar (OpR) rates. OpR rates are available to consumers with PV systems that provide at least 15\% of their annual energy. Under the OpR rates, demand charges are lower than the base TOU rates, while energy charges are higher. The B-19 tariff also offers the Option S for Storage (OpS) rate, which is available to consumers with BES installations that have power ratings of at least 10\% of the consumer's maximum annual demand. The OpS rate imposes low demand charges on daily demand and demand during the typical TOU periods; consumers with sufficient flexibility have the potential to significantly reduce their monthly demand charges. The OpS rate features the same higher energy charges introduced under the OpR rates \cite{e19, b19}.

\section{Mathematical Formulation} \label{Mathematical Formulation}
We define a linear program (LP) to determine the minimum monthly electricity bill, comprised of costs associated with TOU rates and revenues associated with NEM, for a consumer that has a combination of PV and BES assets. Within the LP, BES operation is optimized to achieve the minimum bill, while simulated demand \cite{openei} and PV generation \cite{pvlib} are provided as parameters. To obtain a consumer's total annual electricity bill, the LP is run twelve times. The LP is formulated as follows:
\begin{equation}
    \begin{aligned}
        \min. \quad & D_{max} \cdot DR_{max} + \sum_{p \in \mathcal{P}} D_{tou}(p) \cdot DR_{tou}(p) \\
        & + \frac{1}{4} \cdot \sum_{t \in \mathcal{T}} D_{net}^{+}(t) \cdot ER(t) \\
        & + \frac{1}{4} \cdot \sum_{t \in \mathcal{T}} \left[ D_{net}(t) - D_{net}^{+}(t) \right] \cdot NSR(t)
    \end{aligned}
    \label{eq:obj}
\end{equation}

\noindent subject to:
\begin{equation}
    D_{net}(t) \leq D_{max}, \quad \forall t
    \label{eq:max_dem}
\end{equation}
\begin{equation}
    \delta (t, p) \cdot D_{net}(t) \leq D_{tou}(p), \quad \forall t, \forall p
    \label{eq:tou_dem}
\end{equation}
\begin{equation}
    D_{net}^{+}(t) \geq 0, \quad \forall t
    \label{eq:pos_dem0}
\end{equation}
\begin{equation}
    D_{net}^{+}(t) \geq D_{net}(t), \quad \forall t
    \label{eq:pos_dem_net}
\end{equation}
\begin{equation}
    J(t) = J(t - 1) + \frac{1}{4} \cdot \left[ \eta \cdot P_{cha}(t) - P_{dis}(t) \right], \quad \forall t>0
    \label{eq:soc}
\end{equation}
\begin{equation}
    J(t) = J_{init} + \frac{1}{4} \cdot \left[ \eta \cdot P_{cha}(t) - P_{dis}(t) \right], \quad t=0
    \label{eq:soc0}
\end{equation}
\begin{equation}
    J(T) = J_{init}
    \label{eq:soc_last}
\end{equation}
\begin{equation}
    J_{min} \leq J(t) \leq J_{max}, \quad \forall t
    \label{eq:soc_bounds}
\end{equation}
\begin{equation}
    \frac{1}{4} \cdot \eta \cdot P_{cha}(t) \leq BER - J(t - 1), \quad \forall t>0
    \label{eq:cha_bound}
\end{equation}
\begin{equation}
    \frac{1}{4} \cdot \eta \cdot P_{cha}(t) \leq BER - J_{init}, \quad t=0
    \label{eq:cha_bound0}
\end{equation}
\begin{equation}
    \frac{1}{4} \cdot P_{dis}(t) \leq J(t - 1), \quad \forall t>0
    \label{eq:dis_bound}
\end{equation}
\begin{equation}
    \frac{1}{4} \cdot P_{dis}(t) \leq J_{init}, \quad t=0
    \label{eq:dis_bound0}
\end{equation}
\begin{equation}
    P_{cha}(t) + P_{dis}(t) \leq BPR, \quad \forall t
    \label{eq:power_bound}
\end{equation}
\begin{equation}
    D_{base}(t) + P_{cha}(t) \geq P_{dis}(t), \quad \forall t
    \label{eq:no_export}
\end{equation}

\noindent where
\begin{equation}
    D_{net}(t) = D_{base}(t) - P_{pv}(t) + P_{cha}(t) - P_{dis}(t), \quad \forall t
    \label{eq:net_dem}
\end{equation}

Equation \eqref{eq:obj} is the objective function and reflects the consumer's total electricity bill. The first line of the objective function pertains to a tariff's demand charges, where the first product is the demand charge associated with the maximum monthly demand and the summation of products determines the demand charges for maximum demand during each TOU period. The second line represents the consumer's energy charge and the third line represents the NEM revenue, where the difference represents the consumer's net export to the grid. The NEM sell rate, $NSR$, equals the energy rate minus a non-bypassable charge, which is approximately \$0.02/kWh -- \$0.03/kWh \cite{nem2}. The `$\frac{1}{4}$' multiplier is required because the integration uses a 15-minute time step; this multiplier is included on energy quantities throughout this paper. Constraints \eqref{eq:max_dem} and \eqref{eq:tou_dem} define the maximum demand terms \cite{nguyen2017}. The $\delta$ in Constraint \eqref{eq:tou_dem} is an indicator variable that equals `1' when $t$ aligns with the TOU period, $p$, and `0' otherwise. Constraints \eqref{eq:pos_dem0} and \eqref{eq:pos_dem_net} establish bounds to define the non-negative net demand; net demand is defined in Equation \eqref{eq:net_dem}.

Constraints \eqref{eq:soc} -- \eqref{eq:power_bound} describe the BES model \cite{kirschen_book, nguyen2017}. Constraints \eqref{eq:soc} and \eqref{eq:soc0} describe the state of charge at time $t$. Constraint \eqref{eq:soc_last} ensures that the final state of charge (at $t=T$) equals the initial state of charge. Constraint \eqref{eq:soc_bounds} enforces the upper and lower bounds on the state of charge. Constraints \eqref{eq:cha_bound} and \eqref{eq:cha_bound0} restrict the amount of energy used to charge the BES. Constraints \eqref{eq:dis_bound} and \eqref{eq:dis_bound0} limit the amount of energy that can be discharged from the BES. Constraint \eqref{eq:power_bound} bounds the BES ramping power. Constraint \eqref{eq:no_export} prevents the BES from exporting to the grid, as required by PG\&E's NEM tariff \cite{nem2}.

\section{Case Studies} \label{Case Studies}
We consider two archetypal consumers to assess the impacts of the PG\&E rate redesign: a consumer with morning-and-evening peaking (MEP) demand and a consumer with midday peaking (MDP) demand. The MEP consumer has a maximum demand of 220.9 kW and the MDP consumer has a maximum demand of 326.5 kW. The load profiles for each consumer are adapted from data made available by OpenEI \cite{openei}. Each consumer has a PV system and a two-hour BES. The MEP consumer has a 231.8-kW PV system and a 250-kW BES, while the MDP consumer has a 340.7-kW PV system and a 350-kW BES. Asset sizes are selected so that both consumers have similar ratios of asset size to maximum demand.

The consumers are exposed to each of the PG\&E rates discussed in Section \ref{Overview of PGE Tariffs}:

\begin{itemize}
    \item E-19 base TOU tariff (E19TOU)
    \item E-19 Option R for Solar tariff (E19OpR)
    \item B-19 base TOU tariff (B19TOU)
    \item B-19 Option R for Solar tariff (B19OpR)
    \item B-19 Option S for Storage tariff (B19OpS)
\end{itemize}

\noindent In the following subsections, we examine the impact that these tariffs have on the consumers' total bills and the value added by a BES as the sizes of the consumers' assets are varied.

\subsection{Impact on Total Electricity Bill} \label{Total Bill}

\begin{figure*}
     \centering
     \begin{subfigure}[b]{0.24\textwidth}
         \centering
         \includegraphics[width=\textwidth]{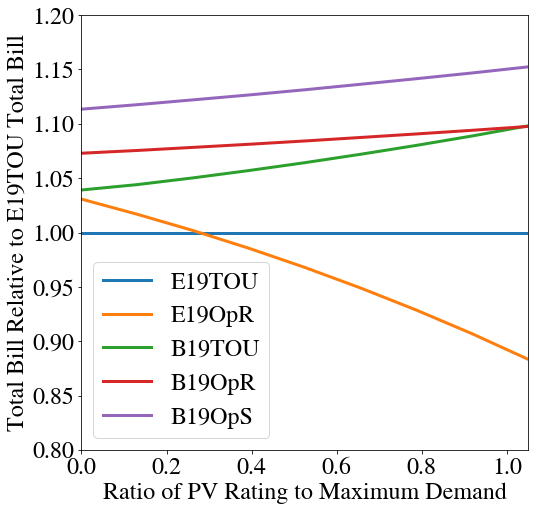}
         \caption{PV capacity sweep (with no BES) for MEP consumer.}
        \label{MEP PV Sweep, no BES}
     \end{subfigure}
     \hfill
     \begin{subfigure}[b]{0.24\textwidth}
         \centering
         \includegraphics[width=\textwidth]{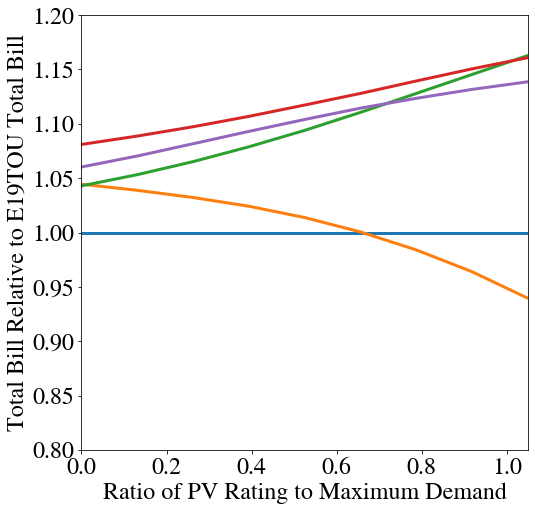}
         \caption{PV capacity sweep (with two-hour BES) for MEP consumer.}
        \label{MEP PV Sweep, with BES}
     \end{subfigure}
     \hfill
     \begin{subfigure}[b]{0.24\textwidth}
         \centering
         \includegraphics[width=\textwidth]{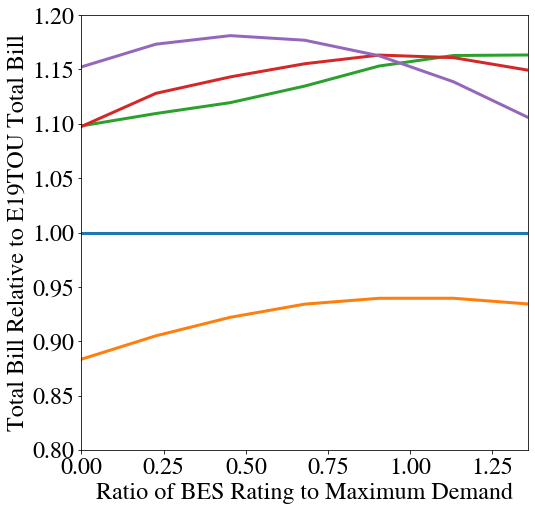}
         \caption{Two-hour BES power capacity sweep for MEP consumer.}
        \label{MEP 2h BES Sweep}
     \end{subfigure}
     \hfill
     \begin{subfigure}[b]{0.24\textwidth}
         \centering
         \includegraphics[width=\textwidth]{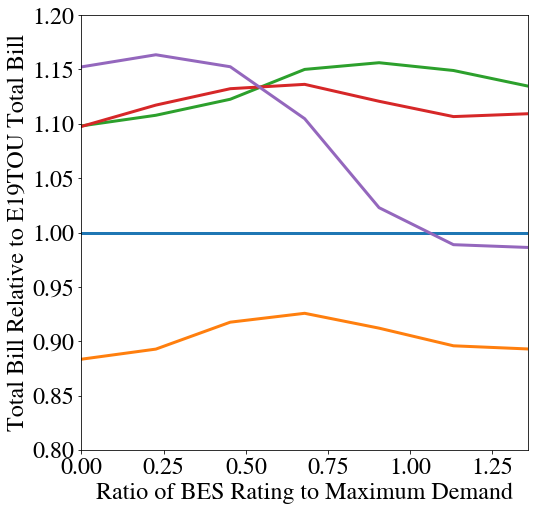}
         \caption{Four-hour BES power capacity sweep for MEP consumer.}
        \label{MEP 4h BES Sweep}
     \end{subfigure}
     \\
     \begin{subfigure}[b]{0.24\textwidth}
         \centering
         \includegraphics[width=\textwidth]{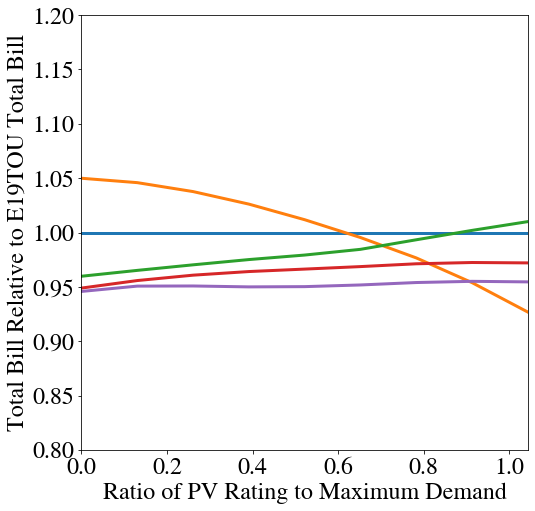}
         \caption{PV capacity sweep (with no BES) for MDP consumer.}
        \label{MDP PV Sweep, no BES}
     \end{subfigure}
     \hfill
     \begin{subfigure}[b]{0.24\textwidth}
         \centering
         \includegraphics[width=\textwidth]{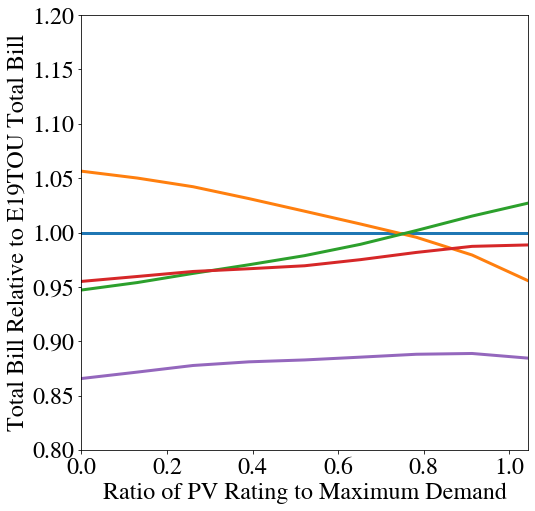}
         \caption{PV capacity sweep (with two-hour BES) for MDP consumer.}
        \label{MDP PV Sweep, with BES}
     \end{subfigure}
     \hfill
     \begin{subfigure}[b]{0.24\textwidth}
         \centering
         \includegraphics[width=\textwidth]{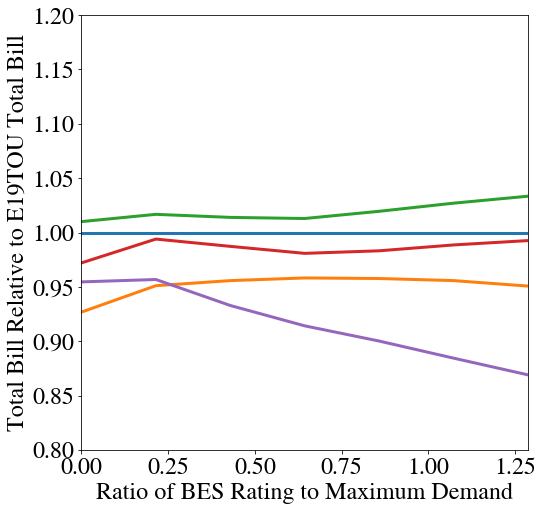}
         \caption{Two-hour BES power capacity sweep for MDP consumer.}
        \label{MDP 2h BES Sweep}
     \end{subfigure}
     \hfill
     \begin{subfigure}[b]{0.24\textwidth}
         \centering
         \includegraphics[width=\textwidth]{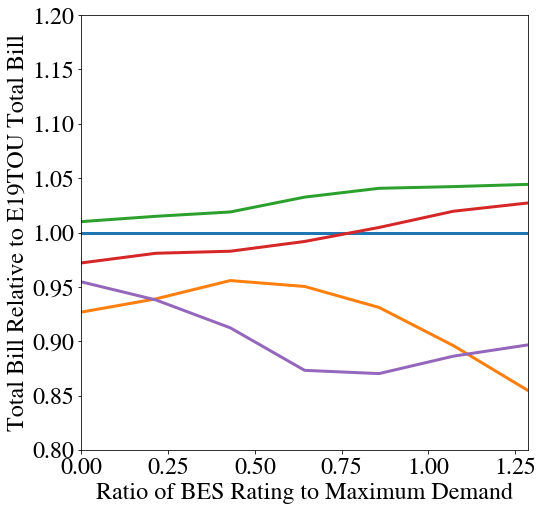}
         \caption{Four-hour BES power capacity sweep for MDP consumer.}
        \label{MDP 4h BES Sweep}
     \end{subfigure}
        \caption{Total bill sensitivity to changes in asset size for two consumer types and five tariffs.}
        \label{fig:tot_bill}
\end{figure*}

As discussed in Section \ref{Overview of PGE Tariffs}, a tariff's TOU periods and demand and energy charges can have a drastic impact on a consumer's total bill. However, the nuances of each tariff can also influence the ability of a consumer's DERs to achieve cost reductions. To understand these interactions, we perform sensitivity analyses for each consumer to examine the sensitivity of their total bill to the size of the assets under each of the five tariffs. We consider four scenarios: 

\begin{enumerate}
    \item a sweep of the consumer's PV system capacity when the consumer does not have a BES,
    \item a sweep of the consumer's PV system capacity when the consumer has the two-hour BES described at the beginning of Section \ref{Case Studies},
    \item a sweep of the consumer's two-hour BES capacity when the consumer has the PV system described at the beginning of Section \ref{Case Studies},
    \item a sweep of a four-hour BES capacity (replacing the consumer's two-hour BES) when the consumer has the PV system described at the beginning of Section \ref{Case Studies}.
\end{enumerate}

\noindent Figure \ref{fig:tot_bill} shows the results of these sensitivity analyses. Figures \ref{MEP PV Sweep, no BES} -- \ref{MEP 4h BES Sweep} show the MEP consumer's results and Figures \ref{MDP PV Sweep, no BES} -- \ref{MDP 4h BES Sweep} show the MDP consumer's results. Since E19TOU is the current base tariff, we present the total bill for each case relative to the total bill under E19TOU.

It is apparent that the demand profile determines how favorable a particular tariff might be for a given consumer. Figures \ref{MEP PV Sweep, no BES} -- \ref{MEP 4h BES Sweep} show that the B-19 tariffs are almost always more expensive for the MEP consumer than the E-19 tariffs, which can be attributed to B-19's later peak and partial-peak demand periods aligning with the MEP consumer's evening demand peaks; under the E-19 rates, the MEP consumer is largely insulated from the peak demand period prices. Conversely, Figures \ref{MDP PV Sweep, no BES} -- \ref{MDP 4h BES Sweep} show that the B-19 tariffs produce lower total bills than the E-19 tariffs for the MDP consumer, particularly for smaller relative asset sizes.

PV system sizing plays a large role in reducing costs for the MEP consumer. Figures \ref{MEP PV Sweep, no BES} and \ref{MEP PV Sweep, with BES} show that E19OpR generally produces lower total bills for the MEP consumer, especially as the PV system size increases. E19OpR, and E19TOU to a lesser extent, outperforms the B-19 rates in large part because of the higher energy rates that coincide with the times of peak solar generation. This allows the MEP consumer to sell excess solar generation through NEM at a higher rate. Consumers with larger PV systems will have more excess power to sell, thereby achieving larger cost reductions. A similar result is not observed when the MEP consumer participates under the B-19 rates because the later peak period  does not coincide with the peak solar generation times. Instead of taking advantage of the higher energy rates to sell excess solar generation, the consumer is simply forced to pay a higher energy charge. PV system size has a much smaller impact on the total bill of the MDP consumer, as evidenced by the insensitive total bills observed in Figures \ref{MDP PV Sweep, no BES} and \ref{MDP PV Sweep, with BES}. The total bill under E19OpR is the noticeable exception, which is to be expected due to the tariff's reliance on selling back excess generation to create cost reductions.

BES sizing appears to have a sizable impact on the total bills of both consumers. For the MEP consumer, Figures \ref{MEP 2h BES Sweep} and \ref{MEP 4h BES Sweep} show that total bills are very sensitive to changes in BES capacity when the consumer is participating under B19OpS. This makes sense due to the multiple smaller demand periods to which the consumer is exposed under B19OpS. B19OpS rewards consumers for having more flexibility, be it through BES with larger capacities or longer durations. Figure \ref{MEP 2h BES Sweep} shows that the larger capacity for the two-hour BES results in a total bill closer to that obtained under E19TOU. As seen in Figure \ref{MEP 4h BES Sweep}, having a BES with a longer duration and a high capacity can result in a total bill that is even lower than that obtained under E19TOU. A similar result is observed for the MDP consumer, where Figures \ref{MDP 2h BES Sweep} and \ref{MDP 4h BES Sweep} show that total bills obtained under B19OpS are most sensitive to changes in BES capacity. The difference is that participating under B19OpS clearly produces the lowest total bill, due to low demand charges and peak solar generation during times of maximum demand. This overlap allows the BES to simultaneously minimize demand charges and increase the amount of excess solar generation that is sold through NEM; BES with larger capacities and a longer duration are best equipped to take advantage of this overlap. It should be noted that Figure \ref{MDP 4h BES Sweep} shows an interesting result, where the four-hour BES with the largest capacity yields the smallest total bill under E19OpR. This likely occurs because the BES is so large that the MDP consumer is able to both sell back large amounts of solar generation, thereby producing large cost reductions through NEM, and adequately shift demand out of peak pricing times.

\subsection{Value Added by a Battery} \label{Battery Value Added}

\begin{figure*}
     \centering
     \begin{subfigure}[b]{0.24\textwidth}
         \centering
         \includegraphics[width=\textwidth]{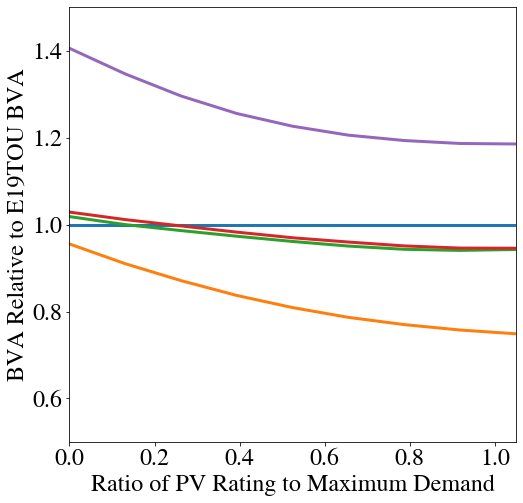}
         \caption{PV capacity sweep (with two-hour BES) for MEP consumer.}
        \label{MEP BVA PV Sweep}
     \end{subfigure}
     \hfill
     \begin{subfigure}[b]{0.24\textwidth}
         \centering
         \includegraphics[width=\textwidth]{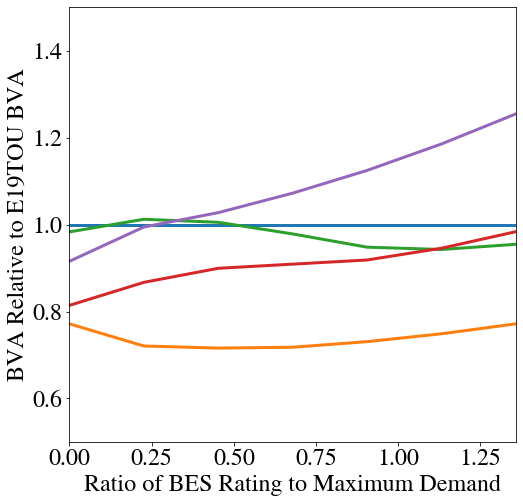}
         \caption{Two-hour BES power capacity sweep for MEP consumer.}
        \label{MEP BVA 2h BES Sweep}
     \end{subfigure}
     \hfill
     \begin{subfigure}[b]{0.24\textwidth}
         \centering
         \includegraphics[width=\textwidth]{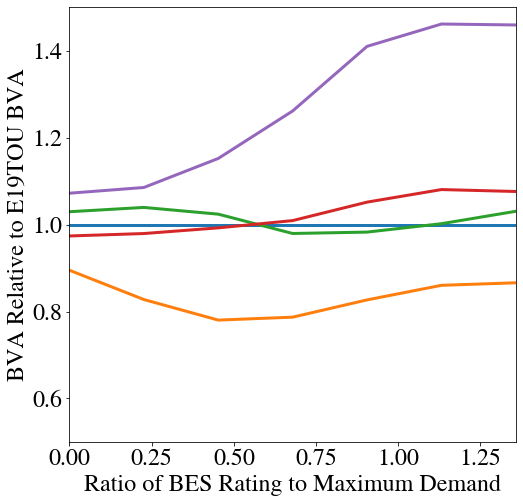}
         \caption{Four-hour BES power capacity sweep for MEP consumer.}
        \label{MEP BVA 4h BES Sweep}
     \end{subfigure}
     \\
     \begin{subfigure}[b]{0.24\textwidth}
         \centering
         \includegraphics[width=\textwidth]{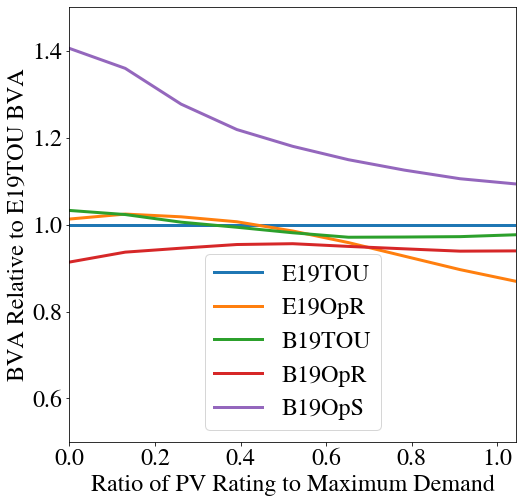}
         \caption{PV capacity sweep (with two-hour BES) for MDP consumer.}
        \label{MDP BVA PV Sweep}
     \end{subfigure}
     \hfill
     \begin{subfigure}[b]{0.24\textwidth}
         \centering
         \includegraphics[width=\textwidth]{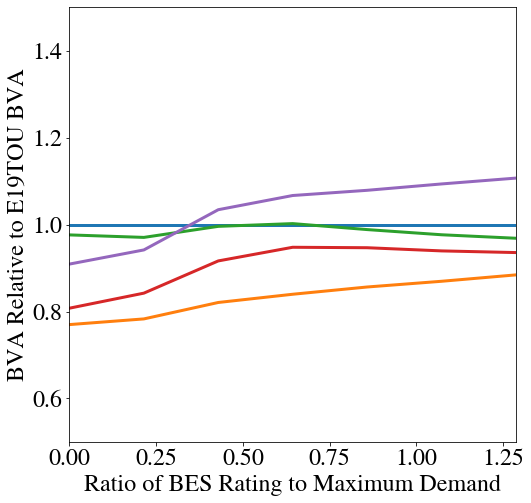}
         \caption{Two-hour BES power capacity sweep for MDP consumer.}
        \label{MDP BVA 2h BES Sweep}
     \end{subfigure}
     \hfill
     \begin{subfigure}[b]{0.24\textwidth}
         \centering
         \includegraphics[width=\textwidth]{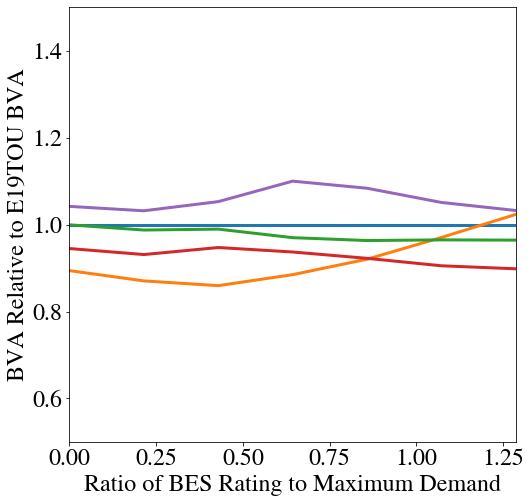}
         \caption{Four-hour BES power capacity sweep for MDP consumer.}
        \label{MDP BVA 4h BES Sweep}
     \end{subfigure}
        \caption{Battery value added (BVA) sensitivity to changes in asset size for two consumer types and five tariffs.}
        \label{fig:bva}
\end{figure*}

As was discussed in Section \ref{Total Bill}, the flexibility provided by a consumer's BES produces similar, if not lower, total bills under the B-19 rates, particularly B19OpS, when compared with the E-19 rates. The apparent need for increased intra-day flexibility will likely make BES a more attractive option to commercial consumers, especially as BES capital costs continue to fall. Knowing this, it is useful to quantify the value that a BES provides to a consumer. Using the results from Section \ref{Total Bill}, we calculate the battery value added (BVA) for each consumer under each tariff. BVA is calculated by subtracting a consumer's total bill with a combination of PV and BES from the same consumer's total bill without BES (but with the same PV capacity). Figure \ref{fig:bva} shows the BVA  for each consumer and a sweep of each asset capacity. Since E19TOU is the current base tariff, we present the BVA results relative to the BVA realized under E19TOU.

As Figure \ref{fig:bva} shows, the tariff under which a consumer takes service has a large impact on the observed BVA. It is clear that consumers taking service under B19OpS generally have the highest BVA, regardless of the consumer's demand profile. This can be largely attributed to B19OpS better rewarding consumer flexibility. Despite E19OpR frequently producing some of the lowest total bills, we see that E19OpR has some of the lowest BVA totals. As previously discussed, E19OpR's reliance on selling excess solar generation to remain cost competitive indicates that the PV system likely has a greater impact on a consumer's cost reductions.

Additionally, the BVA follows similar trends across consumers as asset sizes are varied. Figures \ref{MEP BVA PV Sweep} and \ref{MDP BVA PV Sweep} indicate that BVA performs better relative to E19TOU's BVA as PV capacity decreases. This makes sense because the BES is increasingly relied upon to create cost reductions as the PV system gets smaller. However, this trend does not hold for the MDP consumer under B19OpR, where the relative BVA declines as PV system decreases. Under B19OpR, where the high price period does not coincide with the MDP consumer's peak demand and peak solar generation, declining PV capacity makes it increasingly difficult for the BES to both reduce the consumer's demand peaks and shift demand from the high-priced periods. Figures \ref{MEP BVA 2h BES Sweep} and \ref{MDP BVA 2h BES Sweep} show that BVA relative to E19TOU's BVA generally declines as BES capacity decreases, with the BVA obtained under E19TOU being greatest for two-hour BES with low capacities. Although a similar trend is observed in Figures \ref{MEP BVA 4h BES Sweep} and \ref{MDP BVA 4h BES Sweep} for a four-hour BES, where relative BVA declines with declining BES capacity, we see that the longer duration BES always achieves the greatest BVA under B19OpS, even for the smallest BES capacities.

\section{Conclusions} \label{Conclusions}
This paper shows the impacts that PG\&E's rate redesign can have on consumers with different demand profiles. Without on-site generation or demand flexibility, consumers with demand peaks that are coincident with peak pricing periods will predictably have higher electricity bills. Under PG\&E's rate redesign, it appears that consumers with evening demand peaks stand to be the most adversely affected. However, we see that DERs, particularly BES installations, have the ability to shield consumers from the shifted TOU periods.

While the models used in this paper assume perfect foresight and do not consider BES degradation, we believe the impact of the results are not diminished. The LP described in Section \ref{Mathematical Formulation} does not seek to accurately forecast a consumer's true bill, which is outside the scope of this work. Instead, determining a lower bound on the consumer's bill, where all actions are truly optimal, is sufficient for examining the impacts of PG\&E's rate redesign. BES degradation is not considered because the number of BES cycles averaged close to one per day.

\bibliography{main.bbl}

\begin{thebibliography}{10}
\providecommand{\url}[1]{#1}
\csname url@samestyle\endcsname
\providecommand{\newblock}{\relax}
\providecommand{\bibinfo}[2]{#2}
\providecommand{\BIBentrySTDinterwordspacing}{\spaceskip=0pt\relax}
\providecommand{\BIBentryALTinterwordstretchfactor}{4}
\providecommand{\BIBentryALTinterwordspacing}{\spaceskip=\fontdimen2\font plus
\BIBentryALTinterwordstretchfactor\fontdimen3\font minus
  \fontdimen4\font\relax}
\providecommand{\BIBforeignlanguage}[2]{{%
\expandafter\ifx\csname l@#1\endcsname\relax
\typeout{** WARNING: IEEEtran.bst: No hyphenation pattern has been}%
\typeout{** loaded for the language `#1'. Using the pattern for}%
\typeout{** the default language instead.}%
\else
\language=\csname l@#1\endcsname
\fi
#2}}
\providecommand{\BIBdecl}{\relax}
\BIBdecl

\bibitem{hobbs2019}
B.~F. Hobbs and S.~S. Oren, ``{Three Waves of U.S. Reforms: Following the Path
  of Wholesale Electricity Market Restructuring},'' \emph{IEEE Power \& Energy
  Magazine}, vol.~17, no.~1, pp. 73--81, 2019.

\bibitem{darghouth2014}
N.~R. Darghouth, G.~Barbose, and R.~H. Wiser, ``{Customer-economics of
  residential photovoltaic systems (Part 1): The impact of high renewable
  energy penetrations on electricity bill savings with net metering},''
  \emph{Energy Policy}, vol.~67, pp. 290--300, 2014.

\bibitem{felder2014}
F.~A. Felder and R.~Athawale, ``{The Life and Death of the Utility Death
  Spiral},'' \emph{The Electricity Journal}, vol.~27, no.~6, pp. 9--16, 2014.

\bibitem{b19}
\BIBentryALTinterwordspacing
{Pacific Gas and Electric Company}. (2020) {Electric Schedule B-19: Medium
  General Demand-Metered TOU Service}. Accessed on: 2020-10-27. [Online].
  Available:
  \url{https://www.pge.com/tariffs/assets/pdf/tariffbook/ELEC_SCHEDS_B-19.pdf}
\BIBentrySTDinterwordspacing

\bibitem{darghouth2016}
N.~R. Darghouth, R.~H. Wiser, G.~Barbose, and A.~D. Mills, ``{Net metering and
  market feedback loops: Exploring the impact of retail rate design on
  distributed PV deployment},'' \emph{Applied Energy}, vol. 162, pp. 713--722,
  2016.

\bibitem{darghouth2011}
N.~R. Darghouth, G.~Barbose, and R.~Wiser, ``{The impact of rate design and net
  metering on the bill savings from distributed PV for residential customers in
  California},'' \emph{Energy Policy}, vol.~39, pp. 5243--5253, 2011.

\bibitem{e19}
\BIBentryALTinterwordspacing
{Pacific Gas and Electric Company}. (2020) {Electric Schedule E-19: Medium
  General Demand-Metered TOU Service}. Accessed on: 2020-10-27. [Online].
  Available:
  \url{https://www.pge.com/tariffs/assets/pdf/tariffbook/ELEC_SCHEDS_E-19.pdf}
\BIBentrySTDinterwordspacing

\bibitem{openei}
\BIBentryALTinterwordspacing
E.~Wilson. {``Commercial and Residential Hourly Load Profiles for all TMY3
  Locations in the United States''}. OpenEI. [Online]. Available:
  \url{https://openei.org/doe-opendata/dataset/commercial-and-residential-hourly-load-profiles-for-all-tmy3-locations-in-the-united-states}
\BIBentrySTDinterwordspacing

\bibitem{pvlib}
\BIBentryALTinterwordspacing
{PVLIB Python Community}. {``PVLIB Python Documentation''}. [Online].
  Available: \url{https://pvlib-python.readthedocs.io/en/latest/}
\BIBentrySTDinterwordspacing

\bibitem{nem2}
\BIBentryALTinterwordspacing
{Pacific Gas and Electric Company}. (2020) {Electric Schedule NEM2: Net Energy
  Metering Service}. Accessed on: 2020-10-27. [Online]. Available:
  \url{https://www.pge.com/tariffs/assets/pdf/tariffbook/ELEC_SCHEDS_NEM2.pdf}
\BIBentrySTDinterwordspacing

\bibitem{nguyen2017}
T.~A. Nguyen and R.~H. Byrne, ``{Maximizing the Cost-savings for Time-of-use
  and Net-metering Customers Using Behind-the-meter Energy Storage Systems},''
  in \emph{2017 North America Power Symposium}, 2017.

\bibitem{kirschen_book}
D.~S. Kirschen and G.~Strbac, \emph{{Fundamentals of Power System Economics}},
  2nd~ed.\hskip 1em plus 0.5em minus 0.4em\relax Hoboken, NJ: John Wiley \&
  Sons Ltd, 2019.

\end{thebibliography}

\end{document}